\documentclass[iop]{emulateapj}

\shorttitle{High obliquity orbit for HATS-14b}
\shortauthors{Zhou et al.}

\usepackage{graphicx}
\usepackage{amsmath}
\usepackage{xspace}
\usepackage{array}
{\renewcommand{\arraystretch}{4}%

\newcommand{\myemail}{george.zhou@cfa.harvard.edu}

\begin{document}

\title{A high obliquity orbit for the hot-Jupiter HATS-14b transiting a 5400\,K star}

\author{G.~Zhou\altaffilmark{1,2},
D.~Bayliss\altaffilmark{3},
J.D.~Hartman\altaffilmark{4},
B.J.~Fulton\altaffilmark{5},
G.\'{A}.~Bakos\altaffilmark{4},
A.W.~Howard\altaffilmark{5},
H.~Isaacson\altaffilmark{6},
G.W.~Marcy\altaffilmark{6}, 
B.P.~Schmidt\altaffilmark{2},
R.~Brahm\altaffilmark{7,8}, and
A.~Jord\'an\altaffilmark{7,8},  
}
\altaffiltext{1}{Harvard-Smithsonian Center for Astrophysics, 60 Garden St., Cambridge, MA 02138, USA; \email{\myemail}}
\altaffiltext{2}{Research School of Astronomy and Astrophysics, Australian National University, Cotter Rd, Weston Creek, ACT 2611, Australia}
\altaffiltext{3}{Observatoire Astronomique de l'Universit\'{e} de Gen\`{e}ve, 51 ch. des Maillettes, 1290 Versoix, Switzerland}
\altaffiltext{4}{Department of Astrophysical Sciences, Princeton University, NJ 08544, USA}
\altaffiltext{5}{Institute for Astronomy, University of Hawaii at Manoa, Honolulu, HI, USA}
\altaffiltext{6}{Department of Astronomy, University of California, Berkeley, CA 94720-3411, USA}
\altaffiltext{7}{Instituto de Astrof\'isica, Facultad de F\'isica,
Pontificia Universidad Cat\'olica de Chile, Av.\ Vicu\~na Mackenna
4860, 7820436 Macul, Santiago, Chile}
\altaffiltext{8}{Millennium Institute of Astrophysics, Chile}
\begin{abstract}
We report a spin-orbit misalignment for the hot-Jupiter HATS-14b, measuring a projected orbital obliquity of $|\lambda|= 76_{-5}^{+4}\,^\circ$. HATS-14b orbits a high metallicity, 5400\,K G dwarf in a relatively short period orbit of 2.8 days.  This obliquity was measured via the Rossiter-McLaughlin effect, obtained with observations from Keck-HIRES. The velocities were extracted using a novel technique, optimised for low signal-to-noise spectra, achieving a high precision of $4\,\text{m\,s}^{-1}$ point-to-point scatter. However, we caution that our uncertainties may be underestimated. Due to the low rotational velocity of the star, the detection significance is dependent on the $v\sin i$ prior that is imposed in our modelling. Based on trends observed in the sample of hot Jupiters with obliquity measurements, it has been suggested that these planets modify the spin axes of their host stars, with an efficiency that depends on the stellar type and orbital period of the system.  In this framework, short-period planets around stars with surface convective envelopes, like HATS-14b, are expected to have orbits that are aligned with the spin axes of their host stars.  HATS-14b, however, is a significant outlier from this trend, challenging the effectiveness of the tidal realignment mechanism.
\end{abstract}


\section{Introduction}
\label{sec:introduction}

Orbital obliquity, the angle between the stellar rotation axis and the normal of the orbital plane, probes the migration history of planetary systems. The vast majority of accurate orbital obliquity measurements for planets have come from in-transit spectroscopic observations of the Rossiter-McLaughlin effect \citep[RM,][]{1924ApJ....60...15R,1924ApJ....60...22M}. The transiting planet successively blocks parts of the rotating stellar disk, inducing an apparent radial velocity shift in high-precision spectroscopic observations. To date, the obliquities of 71 planets have been well measured with the RM effect\footnote{Unambiguous measurements, with $\Delta \lambda < 30^\circ$, selected from Ren\'{e} Heller's Holt-Rossiter-McLaughlin Encyclopaedia (\url{www.physics.mcmaster.ca/$\sim$rheller}) - July 2015}. 

Exoplanet systems with a variety of spin obliquity angles have been found, suggesting that many of them have undergone orbital migrations dramatically different from that of the early Solar System. Planets can migrate within the proto-planetary disk via planet-gas disk interactions \citep[e.g.][]{1996Natur.380..606L}, resulting in low orbital obliquities. However, a significant number of planets (23\%) are found to have high obliquities, suggesting a more chaotic, dynamic history. For example, dynamic instability from planet-planet scattering can lead planets into high-eccentricity orbits and a wide range of mutual inclinations \citep[e.g.][]{1996Sci...274..954R,1996Natur.384..619W}. Migration via high-eccentricity, high inclination orbits can also be induced by secular perturbations from companions via Kozai-Lidov cycles \citep[e.g.][]{2003ApJ...589..605W,2007ApJ...669.1298F}. In-disk migration can result in large obliquities if the disk is tilted by a nearby stellar companion \citep[e.g.][]{2010MNRAS.401.1505B,2012Natur.491..418B}.

However, the obliquity we measure today may not be the primordial obliquity of the systems. With a sample size of 28 systems at the time, \citet{2010ApJ...718L.145W} noted that hotter stars with radiative envelopes tend to host planets in a variety of obliquity angles, while the spin-orbit aligned geometry is preferred for cooler stars with convective envelopes. \citet{2010ApJ...719..602S} inferred the line-of-sight spin inclination of 75 planet hosting stars via their rotation periods, finding 10 significantly misaligned systems exclusively orbiting massive stars $(1.2 < M_\star < 1.5\,M_\odot)$. It is thought that the spin direction for the convective envelopes can be modified via tidal interactions with the planet \citep{2012MNRAS.423..486L,2014ApJ...786..102V}. \citet{2012ApJ...757...18A} found that the observed obliquity distribution correlates with the tidal dissipation timescale for each system. Within this framework, large planets in short period orbits around cooler stars should have low obliquities.

The proposed tidal realignment mechanism is not yet well understood. \citet{2012ApJ...757...18A} estimated relative tidal realignment timescales of the planetary systems by calibrating the tidal efficiencies of binary systems hosting radiative and convective stars. To prevent the planets from spirally into the star, \citet{2012MNRAS.423..486L} invoked significantly different stellar tidal $Q$ values governing tidal circularisation and obliquity damping. Models from \citet{2013ApJ...769L..10R} found realignment via tidal dissipation preferentially results in obliquities of 0, 90, or $180^\circ$, the latter two modes are inconsistent with the observations. \citet{2014ApJ...784...66X} found the polar and retrograde modes are unstable, but produces a resulting $\lambda$ too tightly clustered around the prograde mode to replicate the observations. \citet{2014ApJ...790L..31D} argues that the faster magnetic breaking in cooler stars allows the spin axis to be more quickly modified by subsequent tidal interactions.

HATS-14b \citep{2015arXiv150303469M} is a $1.1\,M_\text{Jup}$, $1.0\,R_\text{Jup}$ transiting hot-Jupiter orbiting G7V star with a period of 2.8 days, discovered by the HATSouth survey \citep{2013PASP..125..154B}. The stellar rotation is fast enough that its RM signal is detectable with high precision radial velocity measurements. In this paper, we present in-transit spectroscopy showing that the orbital plane of HATS-14b is significantly misaligned with the stellar rotation axis.

\section{Keck-HIRES observations of the RM effect}
\label{sec:observ-analys}

We observed the spectroscopic transit of HATS-14b on 2015-06-26, from 10:57 -- 14:45 UT, using the High Resolution Echelle Spectrometer \citep[HIRES,][]{1994SPIE.2198..362V} on the Keck-I telescope. A total of 12 observations were made, 7 of which are in transit. The observations span from $\sim 1$ hour before ingress, ending at egress due to the onset of morning twilight. The observations were performed in the standard configuration, with a slit width set to $0.86''$, resulting in a spectral resolution of $\lambda / \Delta \lambda \approx 55000$. The iodine gas absorption cell was used for all the observations. Each exposure was 1200\,s in length. The conditions were clear, and the target remained above airmass 1.8 throughout the observations. 

\begin{figure}
  \centering
  \includegraphics[width=8cm]{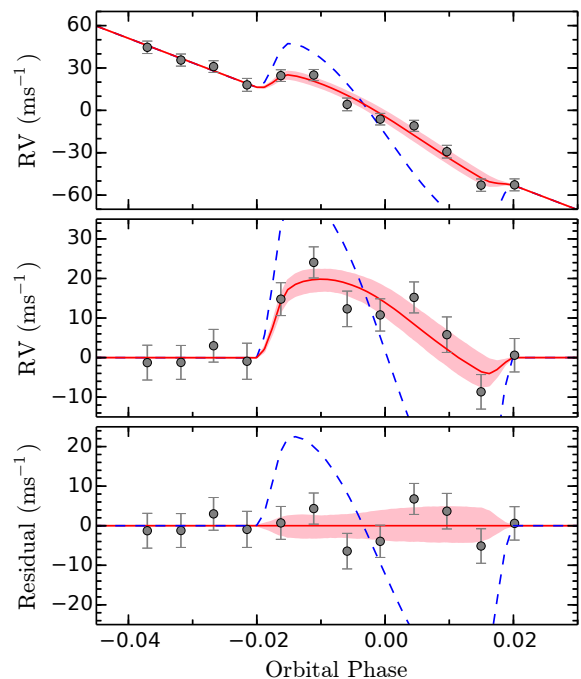}
  \caption{Relative radial velocities from Keck-HIRES for the RM effect of HATS-14b. The red line plots the best fit RM model. The pink region shows the zone where 68\% of the model solutions reside. The expected RM signal from a spin-orbit aligned geometry ($|\lambda| = 0^\circ$), with the same system parameters as HATS-14b, is plotted by the dashed blue line for reference. The top panel plots the observed velocities, the middle panel shows the velocities with the best fit linear trend removed, and the bottom panel shows the velocity residuals from the best fit model.}
  \label{fig:RV}
\end{figure}

Traditionally, radial velocities from iodine-superimposed spectra require an iodine-free spectral template from a separate, higher signal-to-noise observation. At $V_\text{mag} = 13.8$, HATS-14 is a relatively faint target for precise radial velocities. As such, the velocities were derived using a synthetic spectral template, instead of a high signal-to-noise iodine-free observation, based on the techniques developed in \citet{2015arXiv150506738F}. The synthetic template was generated by interpolating spectral models from \citet{2014MNRAS.440.1027C} to the atmospheric parameters of HATS-14 from \citet{2015arXiv150303469M}. The velocities were then measured as per \citet{1996PASP..108..500B}. The synthetic spectrum offers a noise-free template; an equivalent iodine-free observation would have consumed significant telescope time. This novel technique has already delivered high precision multi-epoch radial velocities to enable planet discoveries \citep[e.g. KELT-8b, HATS-8b: ][]{2015arXiv150506738F,2015arXiv150601334B}, but this HATS-14b RM observation offers the first continuous time-series test of the technique. We note the brightness of HATS-14 is similar to that of HATS-8 $(V_\text{mag} = 14.03)$, showing the synthetic template technique consistently works well on fainter targets. The radial velocities are listed in Table~\ref{tab:keck_rvs} and plotted in Figure~\ref{fig:RV}. 

{\renewcommand{\arraystretch}{1.5}
\begin{table}
  \caption{Radial velocities from Keck-HIRES}
  \centering
  \begin{tabular}{lrr}
    \hline\hline
     BJD & RV ($\text{m}\,\text{s}^{-1}$) & $\Delta$RV ($\text{m}\,\text{s}^{-1}$)  \\
    \hline
2457199.95643 & 44.64 & 4.02 \\
2457199.97093 & 35.60 & 3.91 \\
2457199.98501 & 30.99 & 3.77 \\
2457199.99942 & 18.04 & 4.16 \\
2457200.01401 & 24.60 & 3.79 \\
2457200.02819 & 25.00 & 3.59 \\
2457200.04260 & 4.22 & 4.10 \\
2457200.05685 & -6.24 & 3.72 \\
2457200.07145 & -10.96 & 3.58 \\
2457200.08560 & -29.25 & 4.10 \\
2457200.10034 & -52.93 & 3.99 \\
2457200.11478 & -52.73 & 3.86 \\
2457200.12623$^a$ & -20.80 & 5.64 \\
    \hline
  \end{tabular}
  \label{tab:keck_rvs}
\begin{flushleft} 
$^a$ Affected by morning twilight, not used.\\
\end{flushleft}
\end{table}}

\section{Results from global modelling}
\label{sec:global-modelling}

To derive the spin-orbit angle and associated uncertainties, we perform a full global modelling of all available observations for HATS-14b. In addition to the Keck-HIRES velocities, this also includes the observations from \citet{2015arXiv150303469M}: the HATSouth $R$ band discovery light curves, a full transit light curve in $Rc$ band from the 0.3\,m PEST, simultaneous $g$, $r$, $i$, and $z$ band full transit light curves from GROND on the 2.2\,m MPG telescope, and radial velocities of the spectroscopic orbit from FEROS on the 2.2\, MPG telescope, and Coralie on the 1.2\,m Euler telescope. 

The RM effect is modelled using the \emph{ARoME} library \citep{2013A&amp;A...550A..53B}, which provides an analytic model for iodine-derived velocities. The projected spin-orbit angle $\lambda$, line-of-sight stellar rotational velocity $v\sin i$ were free parameters that defined the RM model. A linear trend was also fitted for for the slope of the transit radial velocities, this allows us to account for the effects of velocity jitter on a continuous set of observations, but also removes any constraints the Keck-HIRES velocities have on the spectroscopic orbit \citep{2012ApJ...744..189A}. Following \citet{2013ApJ...772...80F}, we adopted a fixed line broadening of $\beta=3\,\text{km}\,\text{s}^{-1}$ to account for the HIRES instrumental broadening and stellar micro-turbulence effects. We also adopted a fixed macro-turbulence velocity of $\zeta=3.43\,\text{km}\,\text{s}^{-1}$, calculated for HATS-14b as per \citet{2005ApJS..159..141V}. The limb darkening coefficients were fixed to those interpolated from \citet{2000A&amp;A...363.1081C}, calculated for the photometric $V$ band, which corresponds to the iodine-affected region of the observed spectra. In addition, we also allow for a linear trend to the Keck-HIRES velocities in the global fit. Since the RM amplitude is relatively low, we also include the influence of stellar surface convective blueshift in the model as per \citet{2011ApJ...733...30S}. We assume a surface blueshift velocity of $300\,\text{m}\,\text{s}^{-1}$, resulting in a in-transit distortion of $2\,\text{m}\,\text{s}^{-1}$ at maximum. 

We also simultaneously fit for the transit and spectroscopic orbit of HATS-14b. Since the RM signal is dependent on the timing and shape of the transit light curve, this is the only way to ensure the uncertainties and degeneracies in the system parameters are properly propagated. The transit light curves are modelled using a modified version of the JKTEBOP code \citep{1981AJ.....86..102P,2004MNRAS.351.1277S}, and the radial velocities are fitted assuming circular orbits. The free parameters introduced are orbital period $P$, reference transit time $t_0$, planet-star radius ratio $R_p/R_\star$, normalised radius sum $(R_\star+R_p)/a$, line-of-sight inclination $\text{inc}$, and radial velocity orbital semi-amplitude $K$. We also fit for a dilution factor in the HATSouth discovery light curves, to account for biases introduced in the light curve detrending process that can reduce the apparent transit depth. The limb darkening coefficients for the HATSouth discovery and follow-up light curves were fixed to those adopted by \citet{2015arXiv150303469M}. 

The best fit parameters and associated uncertainties are derived using a Markov chain Monte Carlo (MCMC) analysis, using the affine invariant ensemble sampler \emph{emcee} \citep{2013PASP..125..306F}. The per-point uncertainties of each dataset are inflated, where necessary, such that the reduced $\chi^2$ is at unity. This allows us to account for potentially underestimated measurement uncertainties, such as systematic effects, in the observations. The inflation of radial velocity uncertainties is equivalent to adding a jitter term to the radial velocity fit. For the Keck-HIRES velocities, the per-point uncertainties were inflated by a factor of 1.2. A Gaussian prior of $3.1\pm 0.5\,\text{km}\,\text{s}^{-1}$ is applied to $v\sin i$. Uniform priors are assumed for all other parameters. Since the transit geometry is very well constrained by the light curves, the convective blueshift model is fixed throughout the MCMC analysis to that of the best fit geometry, increasing the computation speed.

\begin{figure}
  \centering
  \includegraphics[width=8cm]{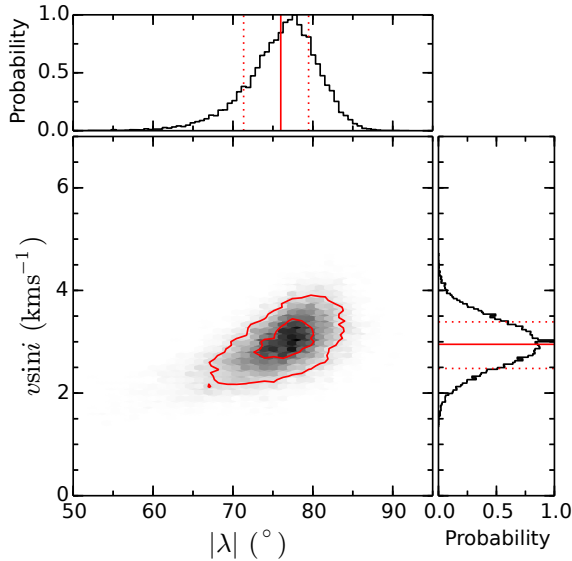}
  \caption{The marginalised posterior probability distribution, showing the correlation between the projected obliquity $|\lambda|$ and the projected stellar rotational velocity $v\sin i$. The 1 and $2\sigma$ confidence regions are marked by the red contours. The marginalised posterior distributions for the $|\lambda|$ and $v\sin i$ parameters are plotted on the side panels, with the median of the distribution marked by the solid line, and the $1\sigma$ confidence region by the dotted lines.}
  \label{fig:posterior}
\end{figure}

We find a best fit misalignment angle of $|\lambda|= 76_{-5}^{+4}\,^\circ$. The full set of parameter solutions are listed in Table~\ref{tab:params}, and the RM model is plotted in Figure~\ref{fig:RV}. The posterior probability distribution for the dependence of $|\lambda|$ on $v\sin i$ is plotted in Figure~\ref{fig:posterior}.

{\renewcommand{\arraystretch}{1.8}
\begin{table}
  \caption{System parameters}
  \centering
  \begin{tabular}{p{4cm}r}
    \hline\hline
    Parameter & Value$^a$ \\
    \hline
    \multicolumn{2}{c}{\emph{Free parameters from global fit}}\\
    Period (days) & $2.766764_{-0.000002}^{+0.000003}$\\
    $t_0$ (BJD) & $2456408.7646_{-0.0002}^{+0.0003}$\\ 
    $(R_\star+R_p)/a$ & $0.126_{-0.001}^{+0.001}$\\
    $R_p/R_\star$ & $0.1143_{-0.0007}^{+0.0007}$\\
    $\text{inc}\,(^\circ)$ & $89.0_{-0.4}^{+0.3}$\\
    HATSouth dilution factor & $0.06_{-0.03}^{+0.03}$\\
    $K \,(\text{m}\,\text{s}^{-1})$ & $158_{-10}^{+11}$\\
    $|\lambda|\,(^\circ)$ & $76_{-5}^{+4}$\\
    $v\sin i \,(\text{km}\,\text{s}^{-1})\,^b$ & $3.0_{-0.5}^{+0.5}$\\
    \multicolumn{2}{c}{\emph{Stellar and planet parameters $^c$}}\\
    $T_\text{eff}$ (K) & $5408\pm65$\\
    $\text{[Fe/H]}$ & $0.28\pm0.03$\\
    $v\sin i\,(\text{km}\,\text{s}^{-1})$ & $3.1\pm0.5$\\
    Isochrone age (Gyr) & $4.9\pm1.7$\\
    $M_\star\,(M_\odot)$ & $0.97\pm0.02$\\
    $R_\star\,(R_\odot)$ & $0.93\pm0.02$\\
    $M_p\,(M_\text{Jup})$ & $1.07\pm0.07$\\
    $R_p\,(R_\text{Jup})$ & $1.04_{-0.02}^{+0.03}$\\
    \hline
  \end{tabular}
\begin{flushleft} 
$^a$ Values are given for the median of the distribution, uncertainties cover the 68\% confidence region.\\
$^b$ A Gaussian prior of $3.1\pm 0.5\,\text{km}\,\text{s}^{-1}$ is applied.\\
$^c$ Selected parameters, updated from \citet{2015arXiv150303469M} with improved analysis of FEROS data as per \citet{2015AJ....150...33B}.\\
\end{flushleft}
  \label{tab:params}
\end{table}}

We note that, due to the slow rotation rate of HATS-14, our largest source of systematic uncertainty in the spin-orbit angle is in the $v\sin i$ prior. The significance of our obliquity detection is a strong function of the $v\sin i$ prior imposed: for a prior of $v\sin i = 2.0 \pm 0.5 \,\text{km}\,\text{s}^{-1}$, the HATS-14b orbit is oblique at $3 \sigma$ significance, while for a prior of $v\sin i = 5.0 \pm 0.5 \,\text{km}\,\text{s}^{-1}$, the orbit is oblique at $19 \sigma$ significance.

To test the influence of the $v\sin i$ Gaussian prior on the derived parameters, we re-run the analysis allowing for a uniform prior on the parameter, deriving $|\lambda|= 77_{-11}^{+7}\,^\circ$ and  $v\sin i = 2.3_{-0.7}^{+2.1}\,\text{km}\,\text{s}^{-1}$; consistent, but larger in uncertainties, with the reported values above. Removing the convective blueshift model in our fit did not bias our results, with $|\lambda|= 76_{-6}^{+4}\,^\circ$ and  $v\sin i = 2.9_{-0.5}^{+0.5}\,\text{km}\,\text{s}^{-1}$. Since the broadening parameters $\beta$ and $\zeta$ affect the shape of the RM model, we also considered their influence on the $|\lambda|$ solution by setting them free, with uniform priors, in the MCMC fit. However, these values are badly constrained and largely degenerate, arriving at $\beta=6_{-3}^{+2}\,\text{km}\,\text{s}^{-1}$ and $\zeta = 3_{-2}^{+3}\,\text{km}\,\text{s}^{-1}$. The resulting solutions for obliquity and projected rotational velocity were consistent with that of the $\beta$ and $\zeta$ fixed models, with $|\lambda| = 73_{-9}^{+6}\,^\circ$ and $v\sin i=3.0_{-0.4}^{+0.4}\,\text{km}\,\text{s}^{-1}$. We also checked for rotational modulation in the HATSouth discovery light curves. A Lomb-Scargle analysis finds a peak at $9.8\pm0.3$ days, with a weak peak-to-peak amplitude of $\sim 3$\,mmag. The same peak is also seen in the autocorrelation function. If this peak is due to rotational modulation, then the period corresponds to a rotational velocity of $4.8 \,\text{km}\,\text{s}^{-1}$, which is consistent within $1\sigma$ with the $v\sin i$. Finally, while the lack of post-egress baseline introduces additional freedom in our global fit, leading to larger reported uncertainties, we caution it may induce additional bias to the $|\lambda|$ measurement.  

\section{Discussions}
\label{sec:discussions}

Using in-transit spectroscopic measurements from Keck-HIRES, we found the hot-Jupiter HATS-14b to be orbiting in a highly inclined plane of $|\lambda|= 76_{-5}^{+4}\,^\circ$.  \citet{2010ApJ...718L.145W} noted an apparent dichotomy in the distribution of planet obliquities, with planets orbiting hotter stars ($T_{\rm eff} < 6250$\,K) having low obliquities, and those around cooler stars with ($T_{\rm eff} > 6250$\,K) having a wide range of obliquities. They hypothesised that this dichotomy may be due to the tidal alignment of the spin axis of the host star to the orbit of the hot Jupiter, with the alignment process being faster for stars with convective envelopes compared to those with radiative envelopes.  This hypothesis was further substantiated by \citet{2012ApJ...757...18A} who presented additional obliquity measurements, and also suggested that a cleaner separation between well-aligned and misaligned systems is found when using the estimated tidal dissipation timescale as a discriminant rather than the stellar effective temperature.  HATS-14 stands out as one of the few G-K dwarfs hosting a short period, misaligned hot Jupiter; it is also one of the few misaligned systems with a tidal timescale below $5 \times 10^{12}$\,yr.

\begin{figure*}
  \centering
{\setlength{\tabcolsep}{0em}
  \begin{tabular}{cc}
    \includegraphics[width=9cm]{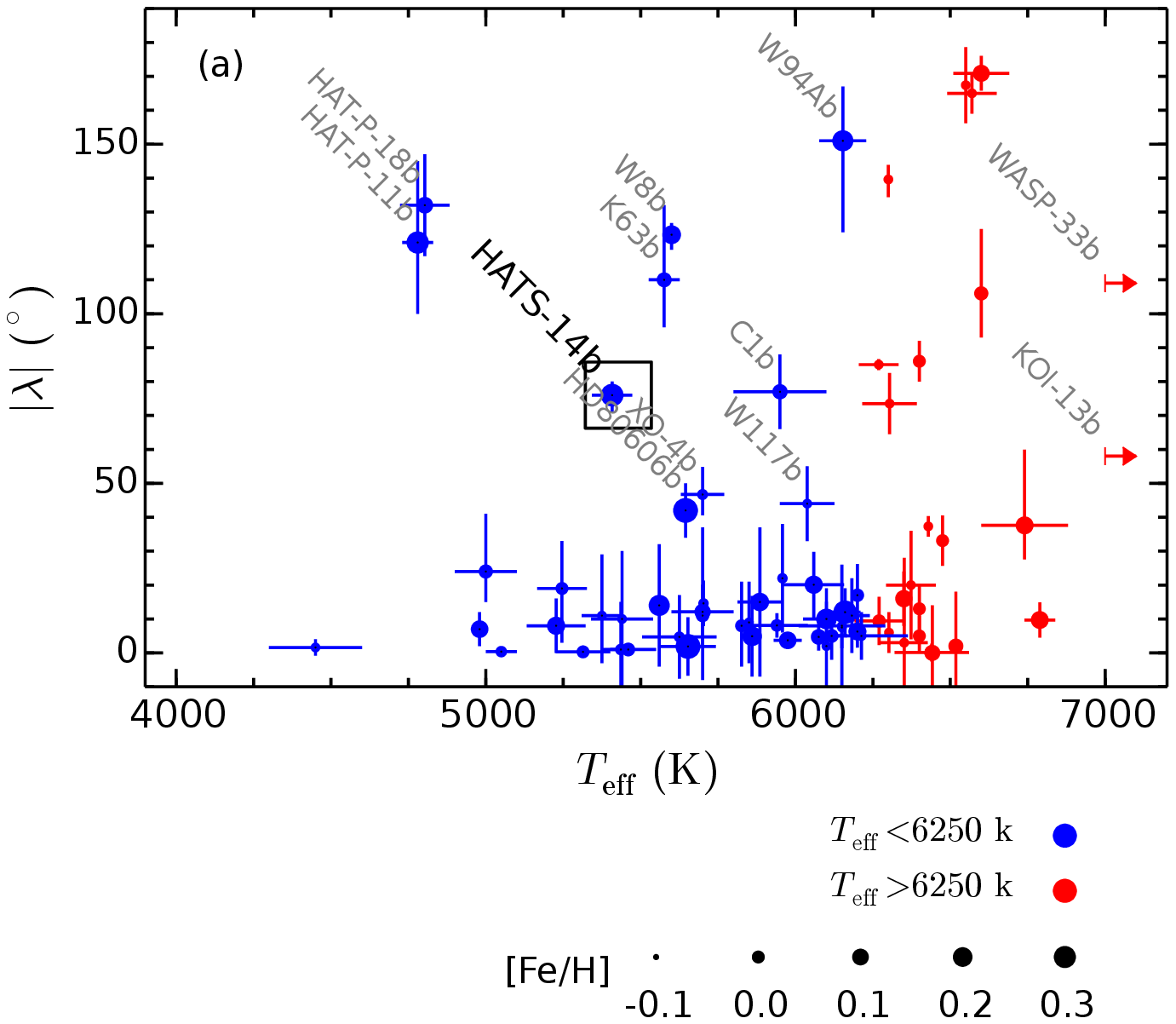}&
    \includegraphics[width=9cm]{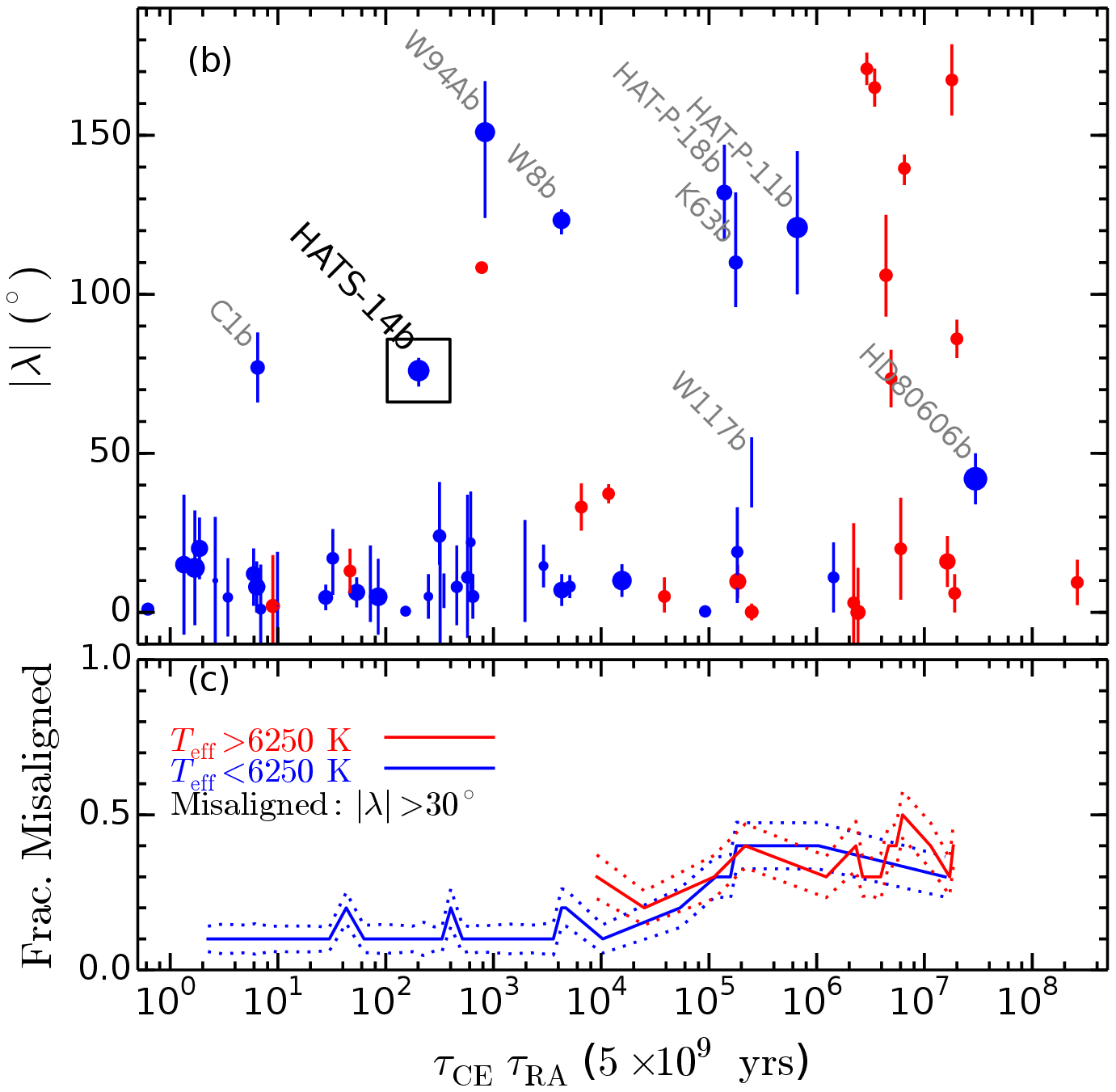}\\
  \end{tabular}}
  \caption{HATS-14b is one of few high obliquity planets orbiting cool stars, and has one of the shortest tidal dissipation timescales. (a): The distribution of projected obliquities $|\lambda|$ as a function of host star effective temperature $T_\text{eff}$ is plotted. Planets with host stars of $T_\text{eff}<6250\,\text{K}$ are plotted in blue, $>6250\,\text{K}$ in red. The marker sizes represent the metallicity of the host stars. Systems around cool host stars with significant misalignments $(|\lambda| > 30^\circ)$ are labelled, with names shortened. `W' represents WASP, `K' for Kepler, `C' for CoRoT. The systems WASP-33b and KOI-13b are hotter than the plot limits, their misalignment angles are plotted and labelled. HATS-14b is also labelled and marked by the black square. (b): Same as (a) but showing the dependence of $|\lambda|$ on the tidal dissipation timescale, calculated from \citet{2012ApJ...757...18A} equation 2 for stars with convective envelopes $(\tau_\text{CE})$, and equation 3 for stars with radiative envelopes $(\tau_\text{RA})$. As per \citet{2012ApJ...757...18A}, a factor of $5\times10^9$ has been removed from the timescales.  (c) The fraction of misaligned planets as a function of $(\tau_\text{CE,RA})$, computed over a moving average over 10 planets per bin, with uncertainties determined from bootstrapping. }
  \label{fig:lambda_distr}
\end{figure*}

To compare HATS-14b to other systems with well measured obliquities, we plot the $|\lambda|$ distribution against host star effective temperature $T_\text{eff}$ and tidal dissipation timescale $\tau$ in Figures~\ref{fig:lambda_distr}a and b\footnote{Assembled from Ren\'{e} Heller's Holt-Rossiter-McLaughlin Encyclopaedia (\url{www.physics.mcmaster.ca/$\sim$rheller}).}. The timescales were calculated as per \citet{2012ApJ...757...18A}. To avoid selection biases, only systems that have been measured via the RM effect are plotted. Systems with shorter dissipation timescales -- those with planets in closer-in orbits and larger planet-star mass ratios -- tend to be spin-orbit aligned. 

Of the measured planets orbiting stars cooler than HATS-14, only the hot-Neptune HAT-P-11b \citep{2010ApJ...710.1724B,2011ApJ...743...61S} and the hot-Saturn HAT-P-18b \citep{2011ApJ...726...52H,2014A&amp;A...564L..13E} exhibit significantly inclined orbital planes. When we consider the tidal dissipation timescales, HATS-14b has a relatively short period orbit ($P\approx 2.8$ days), making its tidal dissipation timescale shorter than most other misaligned systems. In comparison, HAT-P-18 and HAT-P-11 have longer orbital periods and lower masses, resulting in significantly longer dissipation timescales than HATS-14. Only CoRoT-1b, with an orbital period of 1.59 days, exhibits a similar high obliquity at a shorter realignment timescale \citep{2010MNRAS.402L...1P}. Figure~\ref{fig:lambda_distr}c shows the moving average (with bin size of 10 planets) of the fraction of misaligned systems $(|\lambda| > 30^\circ)$ as a function of the tidal dissipation timescale. For systems with timescales like HATS-14b, only $0.10\pm0.05$ are spin-orbit misaligned, and there are none in retrograde orbits. As we tend towards longer dissipation timescales, the misalignment fraction converges towards $0.4\pm0.1$ for $\tau_\text{CE,RA}>5\times10^{14}$, regardless of the host stellar type. 

If tidal realignment is a universal process, it is unclear why planets like HATS-14b and CoRoT-1b remain misaligned, while other planets in the same parameter space were aligned over time. To explain CoRoT-1b,  \citet{2012ApJ...757....6H} pointed out that tidal dissipation acts on the component of the stellar spin parallel to the orbit, making it less effective if the initial alignment is close to polar. Models from \citet{2013ApJ...769L..10R} found realignment via tidal dissipation preferentially results in obliquities of 0, 90, or $180^\circ$. The obliquity of CoRoT-1b $(|\lambda| = 77\pm11^\circ)$ is very similar to that of HATS-14b, and within $\sim 10^\circ$ of the polar geometry, which may contribute to their slower-than-normal realignment. However, only the projected obliquities are known for these systems, making a meaningful assessment difficult. In addition, \citet{2014ApJ...784...66X} found the polar and retrograde modes to be unstable, and \citet{2012ApJ...757...18A} called into question the robustness of the CoRoT-1b measurement. 

High obliquity planets around cool stars, like HATS-14b, may also indicate that tidal realignment is less universal than previously thought. If the observed obliquity distribution is primordial, then we may interpret the temperature-obliquity dichotomy as a dependence of the preferred migration mechanism on stellar mass. The occurrence rate of giant planets increases with stellar mass \citep[e.g.][]{2010PASP..122..905J}, at least until $\sim 2\,M_\odot$ \citep{2015A&amp;A...574A.116R}. Naively, the higher occurrence rate of planets will also make these systems more conducive to dynamic migration, introducing a host star temperature dependence on the obliquity distribution. However, no significant correlation between the eccentricity distribution and stellar mass has been noted, bar the apparent lack of high eccentricity warm-Jupiters around subgiants \citep[e.g.][]{2014A&amp;A...566A.113J}. This suggests that either high-eccentricity migration may not be favourable amongst systems around high mass stars, or that warm-Jupiters around subgiants are subjected to more efficient tidal circularisation and decay \citep[e.g.][]{2013ApJ...772..143S}.

HATS-14 is a particularly metal rich star, with $\text{[Fe/H]} = 0.28\pm0.03$. The metallicity of the host stars are marked in Figure~\ref{fig:lambda_distr}a,b by their point sizes. It has been established that the occurrence rate of giant planets is higher around high-metallicity stars \citep[e.g.][]{2004A&amp;A...415.1153S,2010PASP..122..905J,2012Natur.486..375B}. Giant planet formation via core accretion is more efficient around high metallicity disks, since they facilitate rapid formation of planet cores \citep[e.g.][]{2004ApJ...616..567I}, and have longer dissipation times \citep[e.g.][]{2010MNRAS.402.2735E}. The correlation between metallicity and migrational history is less clear. \citet{2012A&amp;A...541A..97M} found planets in lower metallicity disks migrate further, but form further out compared to higher metallicity disks, negating the effect of metallicity on the observed semi-major axis distribution. \citet{2013ApJ...767L..24D} found eccentric warm-Jupiters preferentially orbit metal rich stars, and proposed planet-planet scattering occurs more frequently around metal rich systems due to the greater abundance of giant planets. If scattering is more efficient for higher metallicity systems, we expect misaligned systems to also be preferentially found around metal-rich stars. However, there is no conclusive difference between the mean host metallicity for aligned systems ($\text{[Fe/H]} = 0.0\pm0.1$) and misaligned systems ($\text{[Fe/H]} = 0.0\pm0.2$), selecting only systems that have not been fully realigned ($\tau_\text{CE,RA}>5\times 10^{11}$ yrs).

\acknowledgments

Work at the Australian
National University is supported by ARC Laureate Fellowship
Grant FL0992131.The authors wish to recognize
and acknowledge the very significant cultural role
and reverence that the summit of Mauna Kea has always
had within the indigenous Hawaiian community. GZ thanks helpful discussions with Simon Albrecht. 
Facilities: \facility{Keck(HIRES)}


\begin{thebibliography}{50}
\expandafter\ifx\csname natexlab\endcsname\relax\def\natexlab#1{#1}\fi

\bibitem[{{Albrecht} {et~al.}(2012{\natexlab{a}}){Albrecht}, {Winn}, {Butler},
  {Crane}, {Shectman}, {Thompson}, {Hirano}, \&
  {Wittenmyer}}]{2012ApJ...744..189A}
{Albrecht}, S., {Winn}, J.~N., {Butler}, R.~P., {et~al.} 2012{\natexlab{a}},
  \apj, 744, 189

\bibitem[{{Albrecht} {et~al.}(2012{\natexlab{b}}){Albrecht}, {Winn}, {Johnson},
  {Howard}, {Marcy}, {Butler}, {Arriagada}, {Crane}, {Shectman}, {Thompson},
  {Hirano}, {Bakos}, \& {Hartman}}]{2012ApJ...757...18A}
{Albrecht}, S., {Winn}, J.~N., {Johnson}, J.~A., {et~al.} 2012{\natexlab{b}},
  \apj, 757, 18

\bibitem[{{Bakos} {et~al.}(2010){Bakos}, {Torres}, {P{\'a}l}, {Hartman},
  {Kov{\'a}cs}, {Noyes}, {Latham}, {Sasselov}, {Sip{\H o}cz}, {Esquerdo},
  {Fischer}, {Johnson}, {Marcy}, {Butler}, {Isaacson}, {Howard}, {Vogt},
  {Kov{\'a}cs}, {Fernandez}, {Mo{\'o}r}, {Stefanik}, {L{\'a}z{\'a}r}, {Papp},
  \& {S{\'a}ri}}]{2010ApJ...710.1724B}
{Bakos}, G.~{\'A}., {Torres}, G., {P{\'a}l}, A., {et~al.} 2010, \apj, 710, 1724

\bibitem[{{Bakos} {et~al.}(2013){Bakos}, {Csubry}, {Penev}, {Bayliss},
  {Jord{\'a}n}, {Afonso}, {Hartman}, {Henning}, {Kov{\'a}cs}, {Noyes},
  {B{\'e}ky}, {Suc}, {Cs{\'a}k}, {Rabus}, {L{\'a}z{\'a}r}, {Papp}, {S{\'a}ri},
  {Conroy}, {Zhou}, {Sackett}, {Schmidt}, {Mancini}, {Sasselov}, \&
  {Ueltzhoeffer}}]{2013PASP..125..154B}
{Bakos}, G.~{\'A}., {Csubry}, Z., {Penev}, K., {et~al.} 2013, \pasp, 125, 154

\bibitem[{{Bate} {et~al.}(2010){Bate}, {Lodato}, \&
  {Pringle}}]{2010MNRAS.401.1505B}
{Bate}, M.~R., {Lodato}, G., \& {Pringle}, J.~E. 2010, \mnras, 401, 1505

\bibitem[{{Batygin}(2012)}]{2012Natur.491..418B}
{Batygin}, K. 2012, \nat, 491, 418

\bibitem[{{Bayliss} {et~al.}(2015){Bayliss}, {Hartman}, {Bakos}, {Penev},
  {Zhou}, {Brahm}, {Rabus}, {Jord{\'a}n}, {Mancini}, {de Val-Borro}, {Bhatti},
  {Espinoza}, {Csubry}, {Howard}, {Fulton}, {Buchhave}, {Henning}, {Schmidt},
  {Ciceri}, {Noyes}, {Isaacson}, {Marcy}, {Suc}, {L{\'a}z{\'a}r}, {Papp}, \&
  {S{\'a}ri}}]{2015arXiv150601334B}
{Bayliss}, D., {Hartman}, J.~D., {Bakos}, G.~{\'A}., {et~al.} 2015, ArXiv
  e-prints

\bibitem[{{Bou{\'e}} {et~al.}(2013){Bou{\'e}}, {Montalto}, {Boisse}, {Oshagh},
  \& {Santos}}]{2013A&amp;A...550A..53B}
{Bou{\'e}}, G., {Montalto}, M., {Boisse}, I., {Oshagh}, M., \& {Santos}, N.~C.
  2013, \aap, 550, A53

\bibitem[{{Brahm} {et~al.}(2015){Brahm}, {Jord{\'a}n}, {Hartman}, {Bakos},
  {Bayliss}, {Penev}, {Zhou}, {Ciceri}, {Rabus}, {Espinoza}, {Mancini}, {de
  Val-Borro}, {Bhatti}, {Sato}, {Tan}, {Csubry}, {Buchhave}, {Henning},
  {Schmidt}, {Suc}, {Noyes}, {Papp}, {L{\'a}z{\'a}r}, \&
  {S{\'a}ri}}]{2015AJ....150...33B}
{Brahm}, R., {Jord{\'a}n}, A., {Hartman}, J.~D., {et~al.} 2015, \aj, 150, 33

\bibitem[{{Buchhave} {et~al.}(2012){Buchhave}, {Latham}, {Johansen},
  {Bizzarro}, {Torres}, {Rowe}, {Batalha}, {Borucki}, {Brugamyer}, {Caldwell},
  {Bryson}, {Ciardi}, {Cochran}, {Endl}, {Esquerdo}, {Ford}, {Geary},
  {Gilliland}, {Hansen}, {Isaacson}, {Laird}, {Lucas}, {Marcy}, {Morse},
  {Robertson}, {Shporer}, {Stefanik}, {Still}, \&
  {Quinn}}]{2012Natur.486..375B}
{Buchhave}, L.~A., {Latham}, D.~W., {Johansen}, A., {et~al.} 2012, \nat, 486,
  375

\bibitem[{{Butler} {et~al.}(1996){Butler}, {Marcy}, {Williams}, {McCarthy},
  {Dosanjh}, \& {Vogt}}]{1996PASP..108..500B}
{Butler}, R.~P., {Marcy}, G.~W., {Williams}, E., {et~al.} 1996, \pasp, 108, 500

\bibitem[{{Claret}(2000)}]{2000A&amp;A...363.1081C}
{Claret}, A. 2000, \aap, 363, 1081

\bibitem[{{Coelho}(2014)}]{2014MNRAS.440.1027C}
{Coelho}, P.~R.~T. 2014, \mnras, 440, 1027

\bibitem[{{Dawson}(2014)}]{2014ApJ...790L..31D}
{Dawson}, R.~I. 2014, \apjl, 790, L31

\bibitem[{{Dawson} \& {Murray-Clay}(2013)}]{2013ApJ...767L..24D}
{Dawson}, R.~I., \& {Murray-Clay}, R.~A. 2013, \apjl, 767, L24

\bibitem[{{Ercolano} \& {Clarke}(2010)}]{2010MNRAS.402.2735E}
{Ercolano}, B., \& {Clarke}, C.~J. 2010, \mnras, 402, 2735

\bibitem[{{Esposito} {et~al.}(2014){Esposito}, {Covino}, {Mancini},
  {Harutyunyan}, {Southworth}, {Biazzo}, {Gandolfi}, {Lanza}, {Barbieri},
  {Bonomo}, {Borsa}, {Claudi}, {Cosentino}, {Desidera}, {Gratton}, {Pagano},
  {Sozzetti}, {Boccato}, {Maggio}, {Micela}, {Molinari}, {Nascimbeni},
  {Piotto}, {Poretti}, \& {Smareglia}}]{2014A&amp;A...564L..13E}
{Esposito}, M., {Covino}, E., {Mancini}, L., {et~al.} 2014, \aap, 564, L13

\bibitem[{{Fabrycky} \& {Tremaine}(2007)}]{2007ApJ...669.1298F}
{Fabrycky}, D., \& {Tremaine}, S. 2007, \apj, 669, 1298

\bibitem[{{Foreman-Mackey} {et~al.}(2013){Foreman-Mackey}, {Hogg}, {Lang}, \&
  {Goodman}}]{2013PASP..125..306F}
{Foreman-Mackey}, D., {Hogg}, D.~W., {Lang}, D., \& {Goodman}, J. 2013, \pasp,
  125, 306

\bibitem[{{Fulton} {et~al.}(2013){Fulton}, {Howard}, {Winn}, {Albrecht},
  {Marcy}, {Crepp}, {Bakos}, {Johnson}, {Hartman}, {Isaacson}, {Knutson}, \&
  {Zhao}}]{2013ApJ...772...80F}
{Fulton}, B.~J., {Howard}, A.~W., {Winn}, J.~N., {et~al.} 2013, \apj, 772, 80

\bibitem[{{Fulton} {et~al.}(2015){Fulton}, {Collins}, {Gaudi}, {Stassun},
  {Pepper}, {Beatty}, {Siverd}, {Penev}, {Howard}, {Baranec}, {Corfini},
  {Eastman}, {Gregorio}, {Law}, {Lund}, {Oberst}, {Penny}, {Riddle},
  {Rodriguez}, {Stevens}, {Zambelli}, {Ziegler}, {Bieryla}, {D`Ago}, {DePoy},
  {Jensen}, {Kielkopf}, {Latham}, {Manner}, {Marshall}, {McLeod}, \&
  {Reed}}]{2015arXiv150506738F}
{Fulton}, B.~J., {Collins}, K.~A., {Gaudi}, B.~S., {et~al.} 2015, ArXiv
  e-prints

\bibitem[{{Hansen}(2012)}]{2012ApJ...757....6H}
{Hansen}, B.~M.~S. 2012, \apj, 757, 6

\bibitem[{{Hartman} {et~al.}(2011){Hartman}, {Bakos}, {Sato}, {Torres},
  {Noyes}, {Latham}, {Kov{\'a}cs}, {Fischer}, {Howard}, {Johnson}, {Marcy},
  {Buchhave}, {F{\"u}resz}, {Perumpilly}, {B{\'e}ky}, {Stefanik}, {Sasselov},
  {Esquerdo}, {Everett}, {Csubry}, {L{\'a}z{\'a}r}, {Papp}, \&
  {S{\'a}ri}}]{2011ApJ...726...52H}
{Hartman}, J.~D., {Bakos}, G.~{\'A}., {Sato}, B., {et~al.} 2011, \apj, 726, 52

\bibitem[{{Ida} \& {Lin}(2004)}]{2004ApJ...616..567I}
{Ida}, S., \& {Lin}, D.~N.~C. 2004, \apj, 616, 567

\bibitem[{{Johnson} {et~al.}(2010){Johnson}, {Aller}, {Howard}, \&
  {Crepp}}]{2010PASP..122..905J}
{Johnson}, J.~A., {Aller}, K.~M., {Howard}, A.~W., \& {Crepp}, J.~R. 2010,
  \pasp, 122, 905

\bibitem[{{Jones} {et~al.}(2014){Jones}, {Jenkins}, {Bluhm}, {Rojo}, \&
  {Melo}}]{2014A&amp;A...566A.113J}
{Jones}, M.~I., {Jenkins}, J.~S., {Bluhm}, P., {Rojo}, P., \& {Melo}, C.~H.~F.
  2014, \aap, 566, A113

\bibitem[{{Lai}(2012)}]{2012MNRAS.423..486L}
{Lai}, D. 2012, \mnras, 423, 486

\bibitem[{{Lin} {et~al.}(1996){Lin}, {Bodenheimer}, \&
  {Richardson}}]{1996Natur.380..606L}
{Lin}, D.~N.~C., {Bodenheimer}, P., \& {Richardson}, D.~C. 1996, \nat, 380, 606

\bibitem[{{Mancini} {et~al.}(2015){Mancini}, {Hartman}, {Penev}, {Bakos},
  {Brahm}, {Ciceri}, {Henning}, {Csubry}, {Bayliss}, {Zhou}, {Rabus}, {de
  Val-Borro}, {Espinoza}, {Jordan}, {Suc}, {Bhatti}, {Schmidt}, {Sato}, {Tan},
  {Wright}, {Tinney}, {Addison}, {Noyes}, {Lazar}, {Papp}, \&
  {Sari}}]{2015arXiv150303469M}
{Mancini}, L., {Hartman}, J.~D., {Penev}, K., {et~al.} 2015, ArXiv e-prints

\bibitem[{{McLaughlin}(1924)}]{1924ApJ....60...22M}
{McLaughlin}, D.~B. 1924, \apj, 60, 22

\bibitem[{{Mordasini} {et~al.}(2012){Mordasini}, {Alibert}, {Benz}, {Klahr}, \&
  {Henning}}]{2012A&amp;A...541A..97M}
{Mordasini}, C., {Alibert}, Y., {Benz}, W., {Klahr}, H., \& {Henning}, T. 2012,
  \aap, 541, A97

\bibitem[{{Pont} {et~al.}(2010){Pont}, {Endl}, {Cochran}, {Barnes}, {Sneden},
  {MacQueen}, {Moutou}, {Aigrain}, {Alonso}, {Baglin}, {Bouchy}, {Deleuil},
  {Fridlund}, {H{\'e}brard}, {Hatzes}, {Mazeh}, \&
  {Shporer}}]{2010MNRAS.402L...1P}
{Pont}, F., {Endl}, M., {Cochran}, W.~D., {et~al.} 2010, \mnras, 402, L1

\bibitem[{{Popper} \& {Etzel}(1981)}]{1981AJ.....86..102P}
{Popper}, D.~M., \& {Etzel}, P.~B. 1981, \aj, 86, 102

\bibitem[{{Rasio} \& {Ford}(1996)}]{1996Sci...274..954R}
{Rasio}, F.~A., \& {Ford}, E.~B. 1996, Science, 274, 954

\bibitem[{{Reffert} {et~al.}(2015){Reffert}, {Bergmann}, {Quirrenbach},
  {Trifonov}, \& {K{\"u}nstler}}]{2015A&amp;A...574A.116R}
{Reffert}, S., {Bergmann}, C., {Quirrenbach}, A., {Trifonov}, T., \&
  {K{\"u}nstler}, A. 2015, \aap, 574, A116

\bibitem[{{Rogers} \& {Lin}(2013)}]{2013ApJ...769L..10R}
{Rogers}, T.~M., \& {Lin}, D.~N.~C. 2013, \apjl, 769, L10

\bibitem[{{Rossiter}(1924)}]{1924ApJ....60...15R}
{Rossiter}, R.~A. 1924, \apj, 60, 15

\bibitem[{{Sanchis-Ojeda} \& {Winn}(2011)}]{2011ApJ...743...61S}
{Sanchis-Ojeda}, R., \& {Winn}, J.~N. 2011, \apj, 743, 61

\bibitem[{{Santos} {et~al.}(2004){Santos}, {Israelian}, \&
  {Mayor}}]{2004A&amp;A...415.1153S}
{Santos}, N.~C., {Israelian}, G., \& {Mayor}, M. 2004, \aap, 415, 1153

\bibitem[{{Schlaufman}(2010)}]{2010ApJ...719..602S}
{Schlaufman}, K.~C. 2010, \apj, 719, 602

\bibitem[{{Schlaufman} \& {Winn}(2013)}]{2013ApJ...772..143S}
{Schlaufman}, K.~C., \& {Winn}, J.~N. 2013, \apj, 772, 143

\bibitem[{{Shporer} \& {Brown}(2011)}]{2011ApJ...733...30S}
{Shporer}, A., \& {Brown}, T. 2011, \apj, 733, 30

\bibitem[{{Southworth} {et~al.}(2004){Southworth}, {Maxted}, \&
  {Smalley}}]{2004MNRAS.351.1277S}
{Southworth}, J., {Maxted}, P.~F.~L., \& {Smalley}, B. 2004, \mnras, 351, 1277

\bibitem[{{Valenti} \& {Fischer}(2005)}]{2005ApJS..159..141V}
{Valenti}, J.~A., \& {Fischer}, D.~A. 2005, \apjs, 159, 141

\bibitem[{{Valsecchi} \& {Rasio}(2014)}]{2014ApJ...786..102V}
{Valsecchi}, F., \& {Rasio}, F.~A. 2014, \apj, 786, 102

\bibitem[{{Vogt} {et~al.}(1994){Vogt}, {Allen}, {Bigelow}, {Bresee}, {Brown},
  {Cantrall}, {Conrad}, {Couture}, {Delaney}, {Epps}, {Hilyard}, {Hilyard},
  {Horn}, {Jern}, {Kanto}, {Keane}, {Kibrick}, {Lewis}, {Osborne},
  {Pardeilhan}, {Pfister}, {Ricketts}, {Robinson}, {Stover}, {Tucker}, {Ward},
  \& {Wei}}]{1994SPIE.2198..362V}
{Vogt}, S.~S., {Allen}, S.~L., {Bigelow}, B.~C., {et~al.} 1994, in Society of
  Photo-Optical Instrumentation Engineers (SPIE) Conference Series, Vol. 2198,
  Instrumentation in Astronomy VIII, ed. D.~L. {Crawford} \& E.~R. {Craine},
  362

\bibitem[{{Weidenschilling} \& {Marzari}(1996)}]{1996Natur.384..619W}
{Weidenschilling}, S.~J., \& {Marzari}, F. 1996, \nat, 384, 619

\bibitem[{{Winn} {et~al.}(2010){Winn}, {Fabrycky}, {Albrecht}, \&
  {Johnson}}]{2010ApJ...718L.145W}
{Winn}, J.~N., {Fabrycky}, D., {Albrecht}, S., \& {Johnson}, J.~A. 2010, \apjl,
  718, L145

\bibitem[{{Wu} \& {Murray}(2003)}]{2003ApJ...589..605W}
{Wu}, Y., \& {Murray}, N. 2003, \apj, 589, 605

\bibitem[{{Xue} {et~al.}(2014){Xue}, {Suto}, {Taruya}, {Hirano}, {Fujii}, \&
  {Masuda}}]{2014ApJ...784...66X}
{Xue}, Y., {Suto}, Y., {Taruya}, A., {et~al.} 2014, \apj, 784, 66

\end{thebibliography}

\end{document}